\documentclass{iopart}

\usepackage{xcolor}
\usepackage{graphicx}
\usepackage{subfigure}

\newcommand{\ket}[1]{\left| #1 \right>} 

\begin{document}
\title{An Integrated Programming and Development Environment for Adiabatic Quantum Optimization}

\date{\today}

\author{T S Humble, A J McCaskey\footnote[1]{Present address: Department of Physics, Virginia Tech, Blacksburg, VA USA}, R S Bennink, J J Billings, \hspace{1cm}E F D'Azevedo, B D Sullivan\footnote[2]{Present address: Department of Computer Science, North Carolina State University, Raleigh, NC USA }, C F Klymko\footnote[3]{Present address: Department of Mathematics and Computer Science, Emory University, Atlanta, GA USA },  H Seddiqi\footnote[4]{Present address: Department of Physics, Georgia Southern University, Statesboro, GA USA}}
\address{Quantum Computing Institute, Oak Ridge National Laboratory, Oak Ridge, TN USA}
\ead{humblets@ornl.gov}
\begin{abstract}
Adiabatic quantum computing is a promising route to the computational power afforded by quantum information processing. The recent availability of adiabatic hardware has raised challenging questions about how to evaluate adiabatic quantum optimization programs. Processor behavior depends on multiple steps to synthesize an adiabatic quantum program, which are each highly tunable. We present an integrated programming and development environment for adiabatic quantum optimization called JADE that provides control over all the steps taken during program synthesis. JADE captures the workflow needed to rigorously specify the adiabatic quantum optimization algorithm while allowing a variety of problem types, programming techniques, and processor configurations. We have also integrated JADE with a quantum simulation engine that enables program profiling using numerical calculation. The computational engine supports plug-ins for simulation methodologies tailored to various metrics and computing resources. We present the design, integration, and deployment of JADE and discuss its potential use for benchmarking adiabatic quantum optimization programs by the quantum computer science community.
\end{abstract}

\maketitle
\setcounter{tocdepth}{2}
\tableofcontents

\section{Introduction}
\label{sec:intro}
The discovery of quantum algorithms with significant speed-ups over their classical counterparts has spurred interest in the research and development of quantum computing systems \cite{Nielsen2000}. Several different but computationally equivalent models for quantum computing have emerged including, in particular, the model of adiabatic quantum computing (AQC) \cite{Farhi2001,Santoro2002}. Notionally, the AQC model for universal quantum computation corresponds to adiabatic (i.e., slow) changes in the state of a quantum physical system. While computationally equivalent to other models, AQC promises some intrinsic benefits for ensuring fault-tolerant computation and reducing system complexity \cite{Childs2001,Young2013,Sarovar2013}.
\par
Additional attention to the AQC model has been stimulated by the recent commercial realization of a special purpose processor that implements the adiabatic quantum optimization (AQO) algorithm \cite{Farhi2000, Farhi2001, Santoro2002}.  The processor, manufactured by the company D-Wave Systems, Inc., realizes a programmable Ising spin-glass model in a transverse field \cite{Harris2010, Berkley2010, Johnson2010, Johnson2011, Bunyk2014}. This hardware is specialized to the AQO algorithm and it is not capable of universal computation within the AQC model, but it does provide a complete realization of a quantum computational device. This has spurred vigorous scientific studies into exactly how the current hardware performs quantum computation, including efforts to differentiate its observed behavior from classical physical processes \cite{Shin2014,Vinci2014}. Moreover, the AQO algorithm is broadly applicable to combinatorial optimization problems and, consequentlt, the D-Wave processor has garnered attention for its potential use in a number of application domains. Examples include problems in classification \cite{Neven2008a,Neven2008b}, machine learning \cite{Pudenz2012}, graph theory \cite{Gaitan2012,Bian2012,Gaitan2013}, artificial neural networks \cite{Neigovzen2009}, and protein folding \cite{PerdomoOrtiz2012} among others \cite{Smelyanskiy2012}.
\par
The availability of quantum hardware allows for benchmarking performance relative to both quantum and classical metrics of computational power. Understanding observed behavior requires a detailed consideration of how the program and hardware interact as well as how the defined metrics represent performance. For example, it is known that performance of the AQO algorithm depends strongly on the specific programming and hardware operation schedules as well as the problem input \cite{Choi2011, Klymko2014}.  Indeed, whereas some studies of the AQO algorithm have reported runtimes that scale polynomially in problem size \cite{Farhi2000b,Farhi2001,Hogg2003,Young2008}, others have suggested worst-case exponential behavior or trapping in local minima \cite{Altshuler2010}. More generally, it has proven difficult to predict the run times of particular problem instances due to the complexity of the underlying quantum dynamics. An essential step in understanding these behaviors is to capture the influence that different programming choices have on observed run times \cite{Farhi2000,Roland2002,Altshuler2010,Dickson2011,Farhi2011,Dickson2012,Boixo2013,McGeoch2013,Katzgraber2014}. 
\begin{figure}[h!]
\centering
\includegraphics[width=0.35\textwidth]{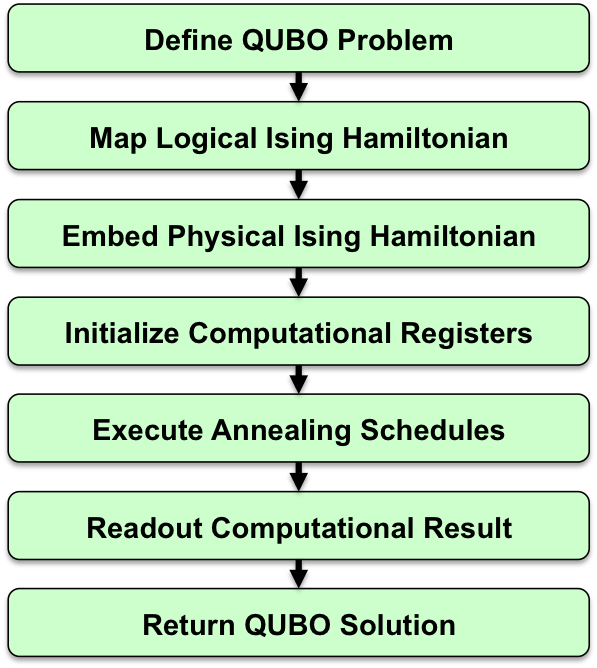}
\caption{A flowchart highlighting the multiple steps taken to synthesize an adiabatic quantum program for the AQO algorithm. The steps are elaborated in detail in Sec.~\ref{sec:aqc}. Briefly, a QUBO problem serves as the classical input to program synthesis while the computed QUBO solution represents the value returned by the program. Each block in the diagram corresponds to a distinct intermediate representation of the quantum program that depends on the choices made in the previous steps.}
\label{fig:flowchart}
\end{figure}
\par
A significant source of the complexity in analyzing implementations of the AQO algorithm arises from the multiple steps undertaken to synthesize the adiabatic quantum program. We provide a brief summary of the process with elaboration of the detailed synthesis deferred to Sec.~\ref{sec:aqc}. Figure ~\ref{fig:flowchart} illustrates that an adiabatic quantum programming process  begins with the reduction of a classical combinatorial optimization problem to a quadratic unconstrained binary optimization (QUBO) problem. The QUBO problem is then mapped into the parameters of an equivalent logical Ising Hamiltonian. The logical Ising Hamiltonian must then be mapped onto the processor as a physical Ising Hamiltonian, a process defined as embedding. This transformation of the reduced problem into a physically realizable program depends on both the hardware layout and the available hardware controls. Ultimately, the computed solution will depend on all previous decisions as well as the actual physics underlying the processor. 
\par
It is currently poorly understood how modifications at the various stages in Fig.~\ref{fig:flowchart} impact the correctness and efficacy of computed solutions. Reconciling the seemingly contradictory results from previous studies as well as understanding more recent experimental benchmarks requires investigating how programming choices impact performance. Motivated by this, we have developed a software environment that captures each step in deriving a program for the adiabatic quantum optimization algorithm. Our framework does not address programming for a universal adiabatic quantum computer, but instead it is specialized to the AQO algorithm and the Ising spin-glass physics underlying the D-Wave processors. The software synthesizes together the steps from Fig.~\ref{fig:flowchart} into an integrated workflow that includes the development of adiabatic quantum programs as well as the collection of diagnostic information for addressing questions about performance.  In the absence of actual hardware, we use numerical simulation to evaluate the variety of programming and operational choices that can effect program behavior. Our simulation capabilities employ multiple numerical methods with the possibility for user extensions. Another important part of the framework is the ability to analyze both the solutions recovered by simulations as well as the intermediate dynamics and Hamiltonians. With the publication of recent benchmarks from available hardware \cite{McGeoch2013,Boixo2013,Ronnow2014}, the ability to make comparisons between simulated and experimental results can be useful for understanding observed behavior. 
\par
The Jade Adiabatic Development Environment (JADE) implements the programming steps highlighted in Fig.~\ref{fig:flowchart}. JADE capabilities include capturing input for a high-level optimization QUBO problem as well as generating the low-level quantum physical program representation. JADE is further integrated with a quantum simulation engine that supports user-defined methodologies for running diagnostic analyses. We present explicit examples of several simulations methodologies based on finite differencing as well as diagnostic derived from the time-dependent eigenspectra and eigenstate populations. 
\par
Because the JADE programming model is tailored to the AQO algorithm and Ising spin glass physics, we suggest that JADE may be useful supporting ongoing benchmark studies of the D-Wave System processor. We do not address the issue of developing benchmarks or methods of evaluating quantum program efficiency, but we do provide a concrete realization of the integrated computational environment needed to carry out such efforts. In particular, our development environment formalizes methods used for programming the quantum processor while offering an interface to simulation for computing detailed diagnostics about how a program executes. For completeness, we note that there is some superficial similarity between JADE and the proprietary BlackBox Compiler from D-Wave Systems, which provides an interface for solving problems on hardware. The primary distinction of JADE is that it enables explicit control of the programming steps for the purpose of testing new programming techniques. Conceptually, a JADE program could be used to drive the actual quantum processor by interfacing with the hardware control system, but we have not explored that option here.
\par
This paper is organized as follows. In Sec.~\ref{sec:aqc}, we summarize the theoretical background leading to Fig.~\ref{fig:flowchart} including the quantum physical basis for AQO. In Sec.~\ref{sec:jademdd}, we present the model-based design of JADE including the system context, implementations of each component, and our test-driven framework for program verification and validation. We present usage results for the case of a recent benchmark problem in Sec.~\ref{sec:results} and we offer conclusions in Sec.~\ref{sec:discussion}. 

\section{Adiabatic Quantum Programming}
\label{sec:aqc}
In this section, we provide a summary of the physical theory and computer science underlying adiabatic quantum programming. This includes the quantum physical description of AQC as well as the steps taken to map the AQO algorithm to a hardware control schedule.
\subsection{Quantum Computational Model}
\label{sec:cm}
The physical basis for the AQC model was first established in terms of quantum annealing by Kadowaki and Nishimori \cite{Kadowaki1998}. Farhi et al.~as well as others later formalized these ideas as a means of universal computation \cite{Farhi2000, Farhi2001,vanDam2001}. Several efforts have since shown the equivalence between the AQC model and other quantum computing models \cite{Aharonov2007, Mizel2007}. In a generalized AQC algorithm \cite{Farhi2000, Farhi2001, Santoro2002}, a quantum physical system of $n$ qubits is evolved under the Schr\"{o}dinger equation
\begin{equation}
\label{eq:tdse}
i \hbar \frac{\partial\ket{\psi(t)}}{\partial t} = H(t) \ket{\psi(t)}
\end{equation}
according to a time-dependent Hamiltonian
\begin{equation}
\label{eq:tdh}
H(t) = A(t) H_{I} + B(t) H_{P}
\end{equation}
that interpolates between the initial Hamiltonian $H_{I}$ and the final (problem) Hamiltonian $H_{P}$ from an initial time $t=t_0$ to a final time $t=T$. We shall assume $t_0 = 0$. In Eq.~(\ref{eq:tdh}), the schedules $A(t)$ and $B(t)$ satisfy the boundary conditions $A(0) \gg B(0)$ and $A(T) \ll B(T)$, while the quantum system is initially prepared in the lowest-energy eigenstate of $H_{I}$. Given the instantaneous eigenvalue equations
\begin{equation}
H(t) \ket{\tilde{\phi}_{j}(t)} = E_{j}(t)\ket{\tilde{\phi}_{j}(t)},
\end{equation}
with $j=0, 1, \ldots {2^{n} - 1}$ labeling states of monotonically increasing energy, the initial state condition implies $\ket{\psi(0)} = \ket{\tilde{\phi}_{0}(0)}$. 
\par
We define the energy gap between the ground and first excited state as
\begin{equation}
\label{eq:gap}
\Delta(t) = E_{1}(t) - E_{0}(t),
\end{equation} 
in which we neglect possible ground state degeneracy for simplicity. If the energy gap is always strictly greater than zero, i.e., $\forall t: \Delta(t) > 0$, then the state $\ket{\psi(t)}$ will remain in the instantaneous ground state with high probability provided certain bounds on the rate of change of the Hamiltonian are satisfied \cite{Farhi2001}. Consequently, evolution under Eq.~(\ref{eq:tdse}) to the time $T$ prepares the final state $\ket{\psi(T)}$ in the lowest energy eigenstate of $H_{P}$. By making a judicious choice of the final Hamiltonian $H_{P}$, the prepared final state may encode the solution to a computation. In order to ensure the computation is correct, the adiabatic condition must be satisfied. In its simplest interpretation, this implies the global time $T$ must be much larger than the inverse of the minimum spectral gap of $H(t)$ \cite{Farhi2001}. More sophisticated analysis shows that better results may be obtained by adjusting the evolution schedule according to the local energy gap \cite{Roland2002}. In either case, failure to ensure the adiabatic condition risks the possibility that the final state will not belong to the ground state manifold of $H_P$ but rather to an excited state and an error in the computation. It is notable that the spectral gap depends not only on the problem to be solved, but also on how the problem is implemented as a quantum program.  Understanding input influence on a program run time and error rates is an open question in quantum computer science.
\subsection{Adiabatic Quantum Optimization}
\label{sec:aqo}
In specializing to the AQO algorithm, we require a quantum logical system of $n$ qubits with an initial Hamiltonian
\begin{equation}
\label{eq:HI}
H_{I} = -\sum_{i\in V_P}{X_{i}}
\end{equation}
and final Ising Hamiltonian
\begin{equation}
\label{eq:ising}
H_{P} = -\sum_{i \in V_P}{\alpha_{i} Z_{i}} - \sum_{(i,j) \in E_P}{\beta_{i,j} Z_{i} Z_{j}},
\end{equation}
where $X_i$ and $Z_i$ are the Pauli operators for the $i$-th qubit \cite{Nielsen2000}, $\alpha_{i}$ is the bias on the $i$-th qubit, and $\beta_{i,j}$ is the coupling between qubits $i$ and $j$. As shown in the section below, the graph $G_{P} = (V_P, E_P)$ with vertex set $V_P$ ($|V_P| \equiv n$) and edge set $E_P$ defines an input optimization problem, cf. the weight matrix $\mathbf{P}$ in Eq.~(\ref{eq:QUBO}). The final Hamiltonian is diagonal in the basis defined by the tensor products of the $\pm 1$ eigenstates of the $Z_i$ operators. This basis will also serve as the computational basis. For comparison, the ground state of the initial Hamiltonian (and initial state of the AQO algorithm) is the symmetric superposition of these computational basis states and has an eigenvalue $-n$.
\par
An important consequence arising from the choice of the initial and final Hamiltonians, respectively, Eqs.~(\ref{eq:HI}) and (~\ref{eq:ising}), is that the time-dependent Hamiltonian $H(t)$ of Eq.~(\ref{eq:tdh}) is not capable of universal adiabatic quantum computation. Extending the form of the Hamiltonian beyond the Ising model, for example, to the 2-local ZZXX Hamiltonian of Biamonte and Love \cite{Biamonte2008}, would support universal computation but we do not consider that case here.
\subsection{Quadratic Unconstrained Binary Optimization}
Any binary optimization problem (BOP) can be mapped into the form of the final Hamiltonian in Eq.~(\ref{eq:ising}). In doing so, we define the classical input to the AQO algorithm as a quadratic unconstrained binary optimization (QUBO) problem. This is because non-binary as well as constrained optimization problems can be reduced to QUBO \cite{Boros2002}, with multiple methods for performing the reduction available \cite{Boros2012}. The QUBO problem is to find
\begin{equation}
\label{eq:QUBO}
\arg\min_{{\bf x}\in {\bf B}^{m}}{{\bf x^{T}}{\bf P} {\bf x}},
\end{equation}
where \textbf{x} is a vector of $n$ binary variables with $x_i \in \{0,1\}$ and \textbf{P} is an $n$-by-$n$ symmetric real-valued cost matrix. We use the weight matrix \textbf{P} to define the weighted adjacency matrix of the input (problem) graph $G_P$ introduced in Eq.~(\ref{eq:ising}). The graph $G_P$ has a vertex set $V_P$ of size $|V_P| \equiv n$ and an edge set $E_{P}$ defined as $(i,j) \in E_P$ iff $P_{i,j} \neq 0$. From this point of view, programming the AQO algorithm requires mapping the matrix $\mathbf{P}$ to the biases and couplings of the Ising Hamiltonian. It has been shown previously by Choi that parameterization of the logical Ising Hamiltonian in Eq.~(\ref{eq:ising}) may be given in terms of the QUBO problem as \cite{Choi2011}
\begin{equation}
\label{eq:ai}
\alpha_{i} = \frac{1}{2} P_{i,i} + \frac{1}{4}\sum_{j=1}^{n}{P_{i,j}}\hspace{1cm}\textrm{for }i = 1\textrm{ to }n,
\end{equation}
and
\begin{equation}
\label{eq:bij}
\beta_{i,j} = \frac{1}{4} P_{i,j}\hspace{1cm}\textrm{for }i < j = 1\textrm{ to }n.
\end{equation}
We may also add an energy shift to the Ising Hamiltonian in Eq.~(\ref{eq:ising}) of the form
\begin{equation}
\gamma = \frac{1}{4}\sum_{i,j=1}^{n}{P_{i,j}} + \frac{1}{2}\sum_{i=1}^{n}{P_{i,i}}
\end{equation}
in order to match the energies of the solution state. Although this shift does not affect the solution obtained using AQC, it must be accounted for in reporting the minimal value in Eq.~(\ref{eq:QUBO}).
\subsection{Hardware Embedding}
Whether or not the logical Hamiltonian in Eq.~(\ref{eq:ising}) is supported directly on a given hardware depends on the available connectivity of that hardware. We express the connectivity of a targeted processor in terms of its hardware graph $G_{H} = (V_H, E_H)$. When any vertex can be coupled to any other vertex and $|V_H| \geq |V_P|$, then it is possible to support all possible input problems using a one-to-one mapping between the logical and physical qubits and the biases and couplings of the physical Hamiltonian. However, when $G_H$ is less than fully connected, then there are certain input problems that will not map directly into hardware. In such circumstances, it may be possible to embed the problem graph $G_P$ into the hardware graph $G_H$ via graph \emph{minor embedding} \cite{Choi2008,Klymko2014}. 
\par
We formally define the {\em minor embedding} of a graph $G_P$ into a graph $G_H$ as a mapping $\phi: V_P \rightarrow V_H$ such that:
\begin{enumerate}
	\item each vertex $i$ in $V_P$ is mapped to the vertex set of a connected subgraph $T_i$ of $G_H$.
	\item if $(i,j) \in E_P$, then there exist $\tau_i, \tau_j \in V_H$ such that $\tau_i \in T_i$, $\tau_j \in T_j$, and $(\tau_i,\tau_j) \in E_H$.
\end{enumerate}
If such a mapping $\phi$ exists, then $G_P$ is {\em minor-embeddable} in $G_H$, or $G_P$ is a {\em minor} of $G_H$. In subsequent discussions, we simply use the term embedding as a reference to minor embedding.
\par
In adiabatic quantum programming, the vertices of the input graph $G_P$ represent the bits of a candidate solution to the QUBO problem, while the edges represent the presence of nonzero coupling coefficients, as defined in Eqs.~(\ref{eq:ai}) and (\ref{eq:bij}), respectively. The vertices of the hardware graph $G_H$ represent the physical qubits and the edges represent the couplings between qubits that are available in the hardware. An embedding maps each vertex in $V_P$ to a subset of $V_H$ and each edge in $E_P$ to edges between these subsets. When an embedding exists, then the resulting subgraph $G^* = (V^*, E^*)$ of the hardware graph defines the physical Ising model
\begin{equation}
\label{eq:hfstar}
H_{G^*} = -\sum_{k \in V^*}{\alpha^*_{k} Z_{k}} - \sum_{(k,\ell) \in E^*}{\beta^*_{k,\ell} Z_{k} Z_{\ell}}
\end{equation}
The bias and coupling coefficients $\alpha^*_k$ and $\beta^*_{k,\ell}$ depend on the selected embedding $\phi$ per the requirements (i) and (ii) listed above. The physical Ising coefficients are defined as \cite{Choi2008}
\begin{equation}
\alpha^{*}_{k} = \alpha_{i} / |T_{i}|\hspace{1cm}\textrm{  for each }k \in V_{T_{i}}
\end{equation}
and for $k \neq \ell$
\begin{equation}
\beta^{*}_{k,\ell} = \left\{
\begin{array}{ll}
\beta_{i,j}/\textit{edges}(T_{i}, T_{j})&\hspace{1cm}\textrm{for }{k \in T_{i}}\textrm{ and }{\ell \in T_{j}}\textrm{ and }{i \neq j} \\
J&\hspace{1cm}\textrm{for }{k \in T_{i}}\textrm{ and }{\ell \in T_{j}}\textrm{ and }{i = j} \\
0&\hspace{1cm}\textrm{otherwise}
\end{array}\right.
\end{equation}
where $\textit{edges}(T_i, T_j)$ is the number of edges between subgraphs $T_{i}$ and $T_{j}$. The constant $J$ is chosen sufficiently large to force the qubits in each subgraph to be strongly correlated, with an upper bound on its value given previously by Choi \cite{Choi2008}. Setting the embedded Ising model coefficients requires knowledge of the matrix $\textbf{P}$ and the selected embedding implied by $G^*$ \cite{Choi2008,Klymko2014}. The embedding need not be unique and, consequently, different instances of the Hamiltonian in Eq.~(\ref{eq:hfstar}) may correspond to the same logical problem of Eq.~(\ref{eq:QUBO}).
\par
\begin{figure}[h]
\centering
\includegraphics[width=2in]{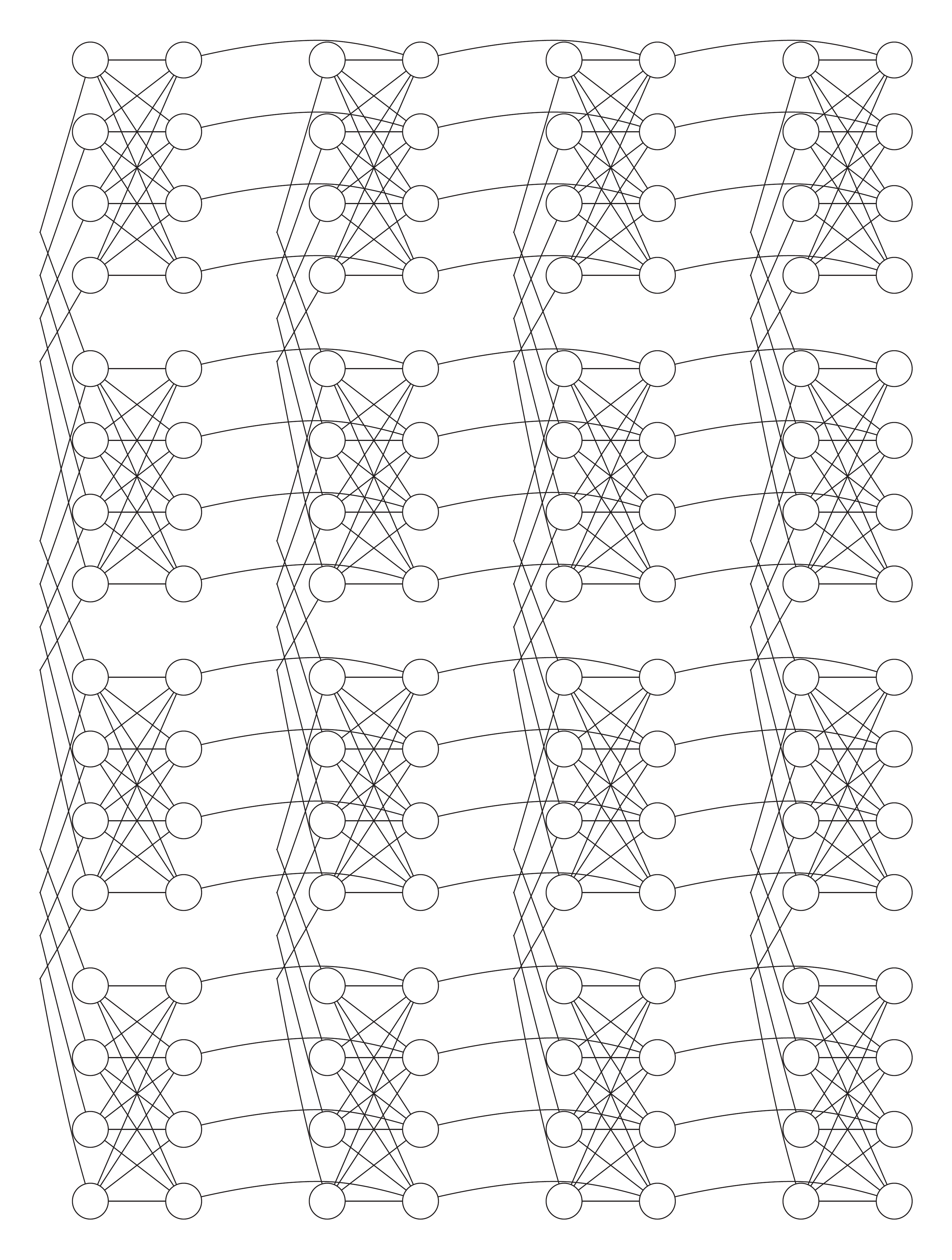}
\caption{A hardware graph for the \emph{Rainier} processor produced by D-Wave Systems, Inc. The design is a $4\times 4$ lattice of interconnected unit cells, with each unit cell is expressed as a $K_{4,4}$ graph. The more recent \emph{Vesuvius} processor has a similar design using an $8\times8$ lattice of unit cells. The geometry of the hardware plays an important role in determining which graphs can be embedded.}
\label{fig:dw1hw}
\end{figure}

A key dependency in finding an embedding is the target hardware graph $G_H$. The hardware graph defines the vertices and connectivity that are available to express the Ising model. An example hardware graph is shown in Fig.~\ref{fig:dw1hw}. Finding those graphs that can be embedded into a fixed hardware graph is an example of subgraph isomorphism, which is known to be NP-Complete in difficulty \cite{Garey1979}. For small hardware graphs, it is tractable to calculate the maximal minors of the graph, i.e., the minors of $G_H$ whose subgraphs represent all other graphs contained in $G_{H}$ \cite{Klymko2014}.  However, this is a brute force approach and therefore does not scale favorably with hardware size. Similarly, attempts to find complete graphs as minor of an arbitrary hardware graph as known to be NP-Complete \cite{Eppstein2009}. Alternatives to these brute force approaches include heuristic algorithms that incorporate knowledge of $G_H$ or that limit the types of input problems \cite{Klymko2014}.
\par
At this point, we emphasize that the role of minor embedding is not as simple as identifying a physical Ising model that is equivalent to the logical Hamiltonian. Indeed, the embedding of a problem into a processor is not unique and, moreover, it is not well understood how different embeddings influence program behavior. There are known tradeoffs in the amount of time spent finding an embedding relative to the size of the embedded problem \cite{Klymko2014}, but it remains unclear how to account for those costs in the benchmarking program performance.
\par
In addition, the current approach to hardware embedding taken by JADE follows the decomposition of a BOP into a QUBO form using quadratization, i.e., decomposing into quadratic form \cite{Boros2012}. However, an alternative programming sequence is to map a BOP directly into a multi-linear Ising model that is then decomposed into bilinear form \cite{Kempe2004}. The latter approach has led to the development of generalized gadgets \cite{Jordan2008} and, more recently, to resource efficient gadgets that replace multi-linear terms in the Hamiltonian with bilinear ones \cite{Biamonte20082, Whitfield2012,Babbush2013}. Gadget decompositions introduce additional ancilla qubits in much the same way that quadratization introduces ancilla bits. Biamonte has presented decompositions minimal in the number of gadget ancilla that would be especially relevant to comparing performance  \cite{Biamonte20082}. We have not explored the use of gadgets in the JADE programming model, or compared the overhead using quadratization, but we believe that the impact of this alternative programming model should be investigated.
\subsection{Hardware Schedules and Program Execution}
We restrict our discussion to AQO algorithms that use a time-dependent Hamiltonian fitting the form of Eq.~(\ref{eq:tdh}), which interpolates between an initial Hamiltonian $H_I$ and the problem Hamiltonian $H_P$ according to the time-dependent annealing schedules $A(t)$ and $B(t)$. More generally, individual biases and couplings can be time-dependent, e.g., $\alpha_{i} = \alpha_{i}(t)$. In either case, the time-dependent schedules specify the rate at which the total Hamiltonian $H(t)$ changes and, consequently, they play an important role in the computational error rates. In particular, the final time $T$ needs to be sufficiently large to ensure the validity of the adiabatic condition, namely,
\begin{equation}
T \gg \frac{\mathcal{E}}{\Delta^{*}}
\end{equation}
where $\Delta^{*} = \min_{t}{\Delta(t)}$ is the global minimum of the spectral gap defined in Eq.~(\ref{eq:gap}) and $\mathcal{E} = \max_t \langle d H(t)/dt \rangle$ is the maximal rate of change during evolution \cite{Farhi2001}. In the absence of information about $\Delta(t)$, it is difficult to ensure the adiabatic condition is satisfied. This uncertainty is one source of the difficulty in benchmarking adiabatic quantum programs. Recent results on amplifying spectral gaps \cite{Somma2013} and developing fault tolerant programs \cite{Lidar2006} suggest new methods for mitigating this uncertainty.
\par
Although the annealing schedules are sufficient for coarsely specifying program execution, it is ultimately necessary to provide the physical implementation of those schedules in terms of hardware controls. The hardware controls that are available for tuning  the biases and coupling of a processor must be capable of expressing programmed schedules. However, available controls are highly dependent on the physics underlying a processor and ensuring the exact implementation of an arbitrary annealing schedule may not be possible. Limitations on annealing schedules arising from constraints and dependencies of control values creates additional uncertainty in the benchmarking effort. Accounting for control constraints and quantification noise is necessary to provide a clear picture of how processor differences impact program behavior. For example, in the case of the family of processors from D-Wave Systems, Inc., biases and couplings can be mapped directly to models for the underlying superconductor Josephson-junction. However, the precision of this mapping is limited by the resolution of the on-board digital to analog converters (DAC's) \cite{Harris2010, Johnson2010}.
\par
In addition to the constraints expected from hardware design, it is also necessary to anticipate the influence of noise on program behavior. Two types of noise affecting quantum dynamics are classical noise in the controls and quantum noise in the system dynamics. Quantum noise may be modeled as an undesired interaction between computational qubits and non-control elements of the hardware. A specific example is the case of thermal influences on the quantum dynamics, which invalidate the pure state description in Sec.~\ref{sec:aqc} and undermine the adiabatic conditions \cite{Dickson2013}. Similarly, classical noise in the hardware controls yields a mixed-state description of the quantum dynamics and may bias program execution away from the solution of interest.
\par
Once the time-dependent behavior of the Hamiltonian $H(t)$ has been fully specified, it remains to execute the program. As noted before, the typical sequence begins by initializing the quantum computational register in the ground state of the initial Hamiltonian $H_I$. How initialization is implemented varies with processor and, more important, it may not be implemented perfectly. This additional source of noise must also be accounted for in evaluating program behavior as it is likely to influence the computational result. The remaining step in  execution is to carry out the hardware control schedule and, therefore, the programmed computation.
\subsection{Computational Readout and Problem Solution}
After evolving to the final time $T$, the state of the computational register is determined using a suitable measurement or readout method. For the case of the AQO algorithm, the ground states at time $T$ are computational eigenstates and, therefore, readout implies a direct measurement in the computational ($Z$) basis. As with program execution, it is more realistic to describe the readout process in terms of the hardware controls. This description includes capturing any noise or uncertainty in the measurement process.
\par
The bit string generated from computational readout is the result of the quantum annealing process. However, mapping this result back to a solution for the original QUBO problem requires decoding measurements according to the inverse of the embedding map. For those cases where a tree of physical qubits represents a single logical qubit, it is necessary to check the value of all such qubits. In cases where measurement results within a tree disagree, then various strategies can resolve the uncertainty. One simple example is to use a majority vote. After decoding the computational readout, a solution to the original QUBO problem is produced and the program is complete. It may be necessary to repeat the execution of the program, for example, to gather statistics on the readout or solution states, however, the steps performed are similar to those described above.

\section{Jade Adiabatic Development Environment}
\label{sec:jademdd}
As presented in Sec.~\ref{sec:aqc}, programming the AQO algorithm for an arbitrary QUBO is a highly tunable process. In this section, we describe a software-based implementation of the process that provides control over each of the programming steps shown in Fig.~\ref{fig:flowchart}. We also describe the integration of this environment with a computational engine that uses numerical simulation for profiling these programs. The simulator is intended for providing insights into how program choices impact program performance.
\par
The Jade Adiabatic Development Environment (JADE) is motivated by the need to provide theoretical benchmarks for current and future adiabatic quantum computing devices. In particular, it was designed to capture insights into the behavior of processor architectures. This is accomplished by using a numerical simulator backend to calculate the time-dependent processor state with respect to programmed algorithm. JADE provides both an engine for simulating the programs that run on adiabatic quantum computing devices and a development environment for specifying program input. In addition, JADE provides methods for constructing adiabatic quantum processor configurations, i.e., the quantum hardware, and for debugging the implementation.
\par
JADE is built using model-driven development, a software development methodology with a strong focus on system use cases as well as architectural extensibility and stability \cite{Nolan2008}. This methodology allows developers to manage system complexity and rigorously verify and validate the final product implementation. Our model-based approach uses the Unified Modeling Language (UML) to capture design decisions and trace requirements \cite{Gamma1995}. We also rely heavily on an object-oriented programming paradigm and software design best practices, such as test driven development \cite{Bittner2003}. 
\subsection{Use Cases}
JADE is designed to provide infrastructure for developing AQO programs and a computational engine for simulating them. This includes functionality for parsing input optimization problems, configuring new quantum hardware, and performing program profiling. Given this broad scope in functionality, JADE was designed for two distinct actors: the \emph{Analyst} and the \emph{Engineer}. 
\begin{figure}[h]
\centering
\includegraphics[width=\textwidth]{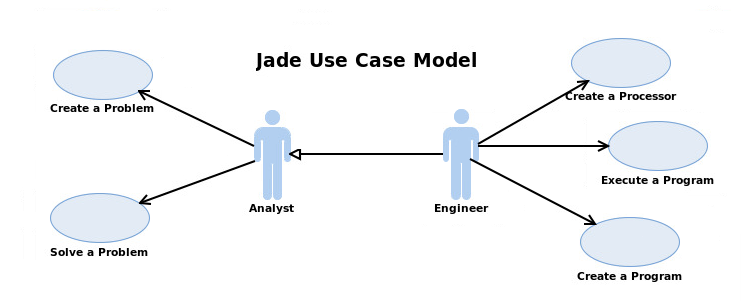}
\caption{The \emph{Analyst} and \emph{Engineer} actors are distinguished by how they use JADE. The \emph{Analyst} is exposed to only a high-level input problem and its computed solution. The \emph{Engineer} has the ability to tune the low-level programming steps and to analyze the computational readout.  The \emph{Engineer} generalizes the \emph{Analyst}, as indicated by the open arrow.}
\label{fig:ucm}
\end{figure}
\par
An \emph{Analyst} represents a JADE user whose primary goal is to solve a discrete optimization problem. The \emph{Analyst} requires a development environment that automates programming choices and execution sequences.  In contrast, an \emph{Engineer} expects to perform additional programming tasks such as customizing low-level Hamiltonian parameters, constructing specialized processor configurations, and defining embedding maps or annealing schedules. As seen in Fig.~\ref{fig:ucm}, this desired JADE functionality is encapsulated by the following use case model:
\begin{itemize}
\item \emph{Create a Problem} - the \emph{Analyst} constructs a discrete optimization problem as either a BOP or QUBO problem. In the case of the former, JADE converts the BOP to its corresponding QUBO representation. This use case creates a \textit{Problem} entity.
\item \emph{Solve a Problem} - the \emph{Analyst} selects a previously created \emph{Problem} to solve using the AQO algorithm. This use case returns a \emph{Solution} entity, which is the computed solution to the input problem. 
\item \emph{Create a Processor} - the \emph{Engineer} creates a processor configuration by specifying the number and connectivity of physical qubits. The \emph{Engineer} may also customize the processor by specifying classical and quantum noise models as well as hardware control constraints. This use case creates a \textit{Processor} entity.  
\item \emph{Create a Program} - the \emph{Engineer} creates a quantum program that is either a logical program or a physical program. A logical program is synthesized from selected \textit{Problem}, \textit{Processor}, and \emph{Embedding} entities, while a physical program is synthesized only from a \textit{Processor}. For the physical program, the \emph{Engineer} sets the parameters of the final Ising Hamiltonian including biases, coupling, and annealing schedules. Both instances of this use case create a \emph{Program} entity.
\item \emph{Execute a Program} - the \emph{Engineer} executes a \emph{Program}. With JADE, the \emph{Engineer} submits the \textit{Program} for simulation along with any profiling and simulations options. This use case creates a \emph{Result} entity that corresponds to the computational readout following program execution. Note that the \emph{Result} of a \emph{Program} does not correspond necessarily with the \emph{Solution} to a \emph{Problem}, as the \emph{Result} may require additional processing to generate a \emph{Solution}.
\end{itemize}
\begin{figure}
\centering
\includegraphics[width=1.0\textwidth]{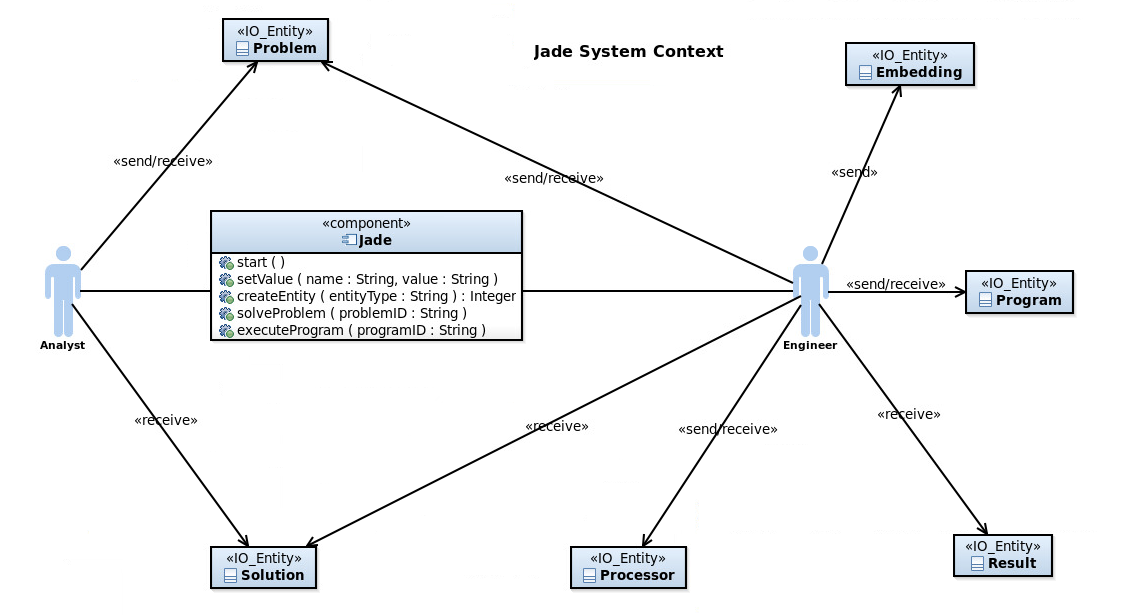}
\caption{The JADE system context represents the interactions between the \emph{Analyst} and \emph{Engineer} actors with the top-level data entities and the software system. Straight lines indicate the assocation between the actors and the system while arrows indicate how the actors interact with the input and output from the system.}
\label{fig:scm}
\end{figure} 
\subsection{System Context}
Alongside the use case model, we also present the system context model in Fig.~\ref{fig:scm}. The system context describes the communication between JADE and its environment as driven by the use case model. The system context details how the \emph{Analyst} and \emph{Engineer} interact with the various input-output (I/O) data. As shown in Fig.~\ref{fig:scm}, the six types of I/O data are: \emph{Problem}, \emph{Processor}, \emph{Embedding},  \emph{Program}, \emph{Result} and \emph{Solution}. These I/O entities are further specified in Sec.~\ref{sec:ca}.
\par
An \emph{Analyst} only has access to \emph{Problem} and \emph{Solution} entities. However, we anticipate that JADE must synthesize other entities internally, for example, a \emph{Program} is required to generate a \emph{Solution}. Consequently, JADE will need private non-interactive methods for internal synthesis of the remaining entities. Although \emph{Processor}, \emph{Embedding}, and \emph{Program} are generated by the system during the \emph{Analyst} workflow, we do not explicitly model that dependency in Fig.~\ref{fig:scm}.
\subsection{Component Architecture}
\label{sec:ca}
JADE comprises three distinct components: \emph{JadeD}, \emph{Sapphire}, and \emph{NiCE}. The \emph{JadeD} component is responsible for data creation, management, synthesis, and verification, i.e., domain logic. The \emph{Sapphire} component is responsible for the simulation of quantum programs according to user-defined plug-ins. The \emph{NiCE} component, a pre-existing open source project, is used to integrate the \emph{JadeD} and \emph{Sapphire} components and to manage the computational work flow \cite{WWWNICE}. Each component provides an independent API.
\begin{figure}[ht]
\begin{center}
\includegraphics[width=\textwidth]{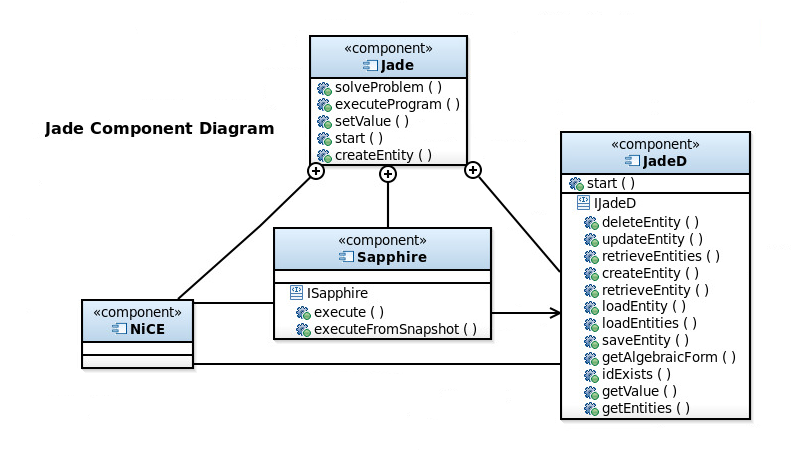}
\end{center}
\caption{JADE comprises three components: \emph{JadeD}, \emph{NiCE}, and \emph{Sapphire}. The interfaces presented for each component are used to manage component interactions and maintain the separation of concerns between domain logic (\emph{JadeD}), workflow management (\emph{NiCE}), and numerical simulation (\emph{Sapphire}). The arrow indicates the uni-directional dependency of \emph{Sapphire} on \emph{JadeD}, while the straightline between \emph{JadeD} and \emph{NiCE} indicates a bi-directional association.}
\label{fig:ca}
\end{figure}
\par
Figure \ref{fig:ca} highlights the interactions between the three components and the associated interfaces. Both \emph{JadeD} and \emph{Sapphire} couple to \emph{NiCE}, which provides a user-driven coupling between program development and program execution. There is a dependency between \emph{JadeD} and \emph{Sapphire} due the latter's need to parse \emph{Program} data structures. This dependency is restricted to a very narrow subset of the \emph{JadeD} functionality and we expect future versions will isolate it in a separate shared library.
\subsection{JadeD}
\label{sec:jaded}
The \emph{JadeD} component handles creation and manipulation of quantum programming by exposing a basic create, retrieve, update, and delete interface. This interface enables generation, manipulation, and persistence of \emph{Entity} data objects, which represent high-level abstractions of the various types of I/O data. The functional scope of \emph{JadeD} includes parsing user-provided input into verified formats, validating that input, and generating subclasses of \emph{Entity} tailored to specific input types. We define an \emph{IJadeD} interface to specify how the \emph{JadeD} component interacts with clients. By defining a formal interface, we are able to offer the option of supporting multiple \emph{JadeD} variants.  
\begin{figure}[ht]
\begin{center}
\includegraphics[width=5.0in]{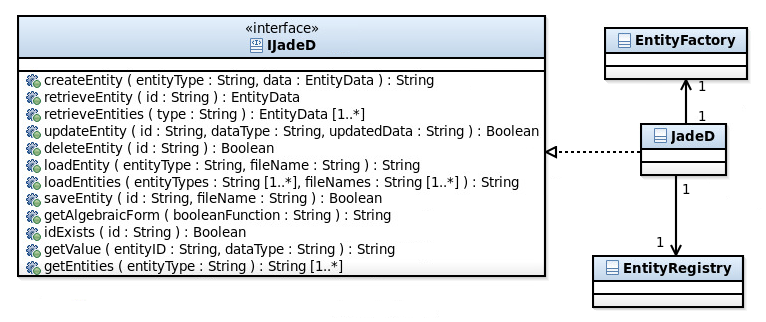}
\end{center}
\caption{The \emph{IJadeD} interface defines the methods exposed to the user or external application. \emph{JadeD} implements this interface by making use of various data entities that can be generated using a factory pattern and managed using a common registry.}
\label{fig:jadedwb}
\end{figure}
\par
As shown in Fig.~\ref{fig:jadedwb}, the \emph{IJadeD} interface includes a number of methods for creating and storing entity instances. The \emph{JadeD} class is a realization of this interface that provides a concrete implementation of the defined functionality.  The \emph{JadeD} implementation presented here uses a variety of object-oriented design patterns with the factory design pattern being the most significant \cite{Gamma1995}. The factory pattern is used to create and modify entities in an abstract manner, which pushes the underlying details of construction to the varying entity subclasses. A registry enhances this factory pattern by permitting the sharing of objects across domain boundaries. The use of factories and a data registry allows future developers to add new entity specializations in an easy and efficient manner. In Fig.~\ref{fig:jadedwb}, the factory pattern and the corresponding data registry are implemented as \emph{EntityFactory} and \emph{EntityRegistry}, respectively. 
\subsubsection{Graph}
\begin{figure}[h]
\begin{center}
\includegraphics[width=5.0in]{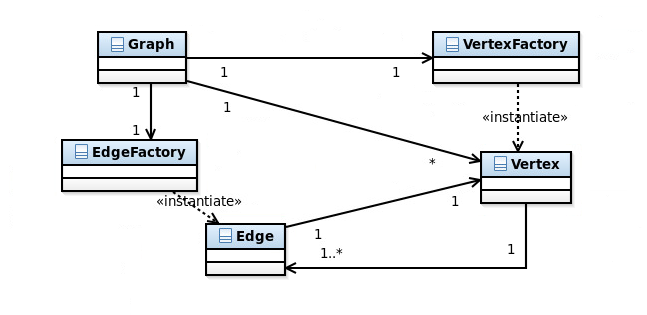}
\end{center}
\caption{The \emph{Graph} class encapsulates vertices and edges, whose respective implementations use the \emph{VertexFactory} and \emph{EdgeFactory} factory patterns.  Open arrows indicate the source associates with the target class while multiplicities annotating the arrowhead identify the number of instance ($*$ denotes unlimited instances).}
\label{fig:graph}
\end{figure}

The \emph{Graph} data structure represents a set of vertices together with a set of edges coupling those vertices. Graph structures are common to the \emph{Problem}, \emph{Processor}, \emph{Embedding}, and \emph{Program} entities. The \emph{JadeD} \emph{Graph} model shown in Fig.~\ref{fig:graph} provides an abstraction of this structure in a way that promotes customization and extensibility with respect to a given entity type. 
\par
In supporting this versatility, the \emph{Graph} class utilizes two factory design patterns for generating vertices and edges \cite{Nolan2008}. This ensures object polymorphism by allowing custom subclasses to inject specialized edges and vertices. For example, this mechanism allows the production of static graphs for \emph{Problem}, graphs that evolve in time for \emph{Program}, and graphs that alter their state according to predefined conditions or controls for \emph{Processor}. 
\subsubsection{Problem} 
The \emph{Problem} class is a subclass of \emph{Entity} that encapsulates the input data describing a discrete optimization problem. It is created by either an \emph{Analyst} or \emph{Engineer} in order to define the logical problem that the system will solve.
\par
The current implementation of \emph{JadeD} permits users to construct two distinct types of \emph{Problem}. The first is a weighted or pseudo-Boolean optimization problem. The user inputs an arbitrary number of Boolean clauses in terms of the literals $b_i$, e.g., $\left((b_{1} \textsc{ AND } b_{2}) \textsc{ OR} \textsc{ NOT } b_3\right)$, and each clause also has an associated real-valued weight $w_i$. The pseudo-Boolean function is then cast into an equivalent BOP by converting each Boolean literal to a corresponding binary variable, e.g., $b_i \mapsto x_i$, \textsc{True} $\mapsto 1$ and \textsc{False} $\mapsto 0$. The Boolean clauses are then recast into equivalent binary arithmetic expressions. Denoting the $i$-th binary arithmetic clause as $f_i$ and the corresponding weight as $w_i$, the equivalent BOP over $m$ bits is 
\begin{equation}
\label{eq:bop}
\arg\min_{\mathbf{x}}\sum\limits_{i} w_i f_i(\mathbf{x}),
\end{equation}
where $\textbf{x} \in \{0,1\}^{m}$ is an $m$-bit vector \cite{Boros2002}. In JADE, the \emph{BOP} class stores both the original Boolean clauses and the reductions to algebraic expressions with corresponding weights.
\par
\begin{figure}[ht]
\begin{center}
\includegraphics[width=\textwidth]{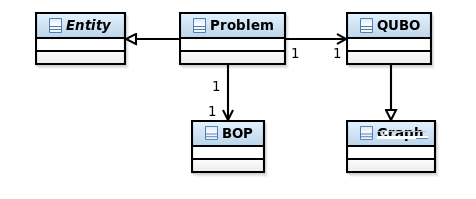}
\end{center}
\caption{The dependencies of \emph{Problem} on \emph{BOP} and \emph{QUBO} entities. \emph{Problem} generates \emph{QUBO} from an input \emph{BOP}. Alternatively, the \emph{QUBO} may be supplied directly.}
\label{fig:problem}
\end{figure}

The second type of \emph{Problem} supported by \emph{JadeD} is the QUBO problem defined in Eq.~(\ref{eq:QUBO}). For this type, the input corresponds to the elements of the matrix $\mathbf{P}$. The matrix $\mathbf{P}$ is then interpreted as a weighted adjacency matrix and parsed by \emph{JadeD} into a \emph{Graph}. Accordingly, the \emph{QUBO} class is a subclass of \emph{Graph}. The dependencies between the various \emph{Problem} subclasses are illustrated in Fig.~\ref{fig:problem}.
\par
As discussed in Sec.~\ref{sec:aqc}, a BOP of the form in Eq.~(\ref{eq:bop}) can be reduced to a corresponding QUBO problem of the form in Eq.~(\ref{eq:QUBO}). The reduction, however, requires introduction of penalty terms to replace multilinear terms with quadratic or linear terms \cite{Boros2002}. Expressing these penalties ultimately requires additional ancilla bits which enlarge the binary state space. When \emph{JadeD} instantiates a \emph{BOP}, the corresponding \emph{QUBO} is immediately generated as part of the \emph{Problem}. \emph{JadeD} uses a QUBO reduction method that replaces the product of two binary variables by a new binary variables; the process repeats until a quadratic form remains \cite{Boros2012}. The relevant \emph{BOP} information is maintained as part of the \emph{Problem} in order to facilitate developing the \emph{Solution} entity returned to the \emph{Analyst}.
\subsubsection{Processor}
The \emph{Processor} entity encapsulates the structure and behavior of a quantum hardware configuration. It generalizes \emph{Graph} by using an adjacency matrix with unit diagonal entries to indicate vertex availability and unit off-diagonal entities for available connections between qubits. \emph{Processor} wraps a subclass of \emph{Graph} referred to as \emph{Hardware} and provides methods to query and manipulate its structure. The \emph{Hardware} subclass can also implement the embedding of an input \emph{Problem} into the hardware. This produces an \emph{Embedding} entity, which subclasses \emph{Graph} to express the graph $G^{*}$ that defines the embedded Hamiltonian $H_{G^*}$ from Eq.~(\ref{eq:hfstar}). 
\par
\emph{Processor} also allows users to specify a functional time dependence for the bias and coupling parameters of vertices. The \emph{Control} class encapsulates functions to express the Ising model parameters in terms of physical quantities that directly influence hardware behavior. For the example of a D-Wave processor, the parameters of the Ising Hamiltonians are mapped into the bias and tunneling energies of the superconducting flux qubits \cite{Harris2010}. These physical quantities are controlled experimentally in terms of the applied current and magnetic flux, and the \emph{Control} class allows the developer to express this dependency. Custom noise models for these controls can also be added to \emph{Processor} through the \emph{Noise} class, which can express both classical and quantum noise functions. 
\begin{figure}[ht]
\begin{center}
\includegraphics[width=4.0in]{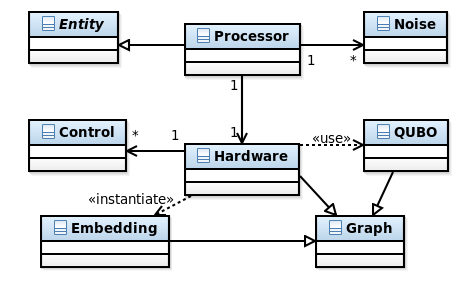}
\end{center}
\caption{The dependencies of the \emph{Processor} class, which includes \emph{Hardware}, \emph{Noise}, and \emph{Control} entities. The \emph{Embedding} entity is instantiated after a QUBO is embedded into the \emph{Processor}.}
\label{fig:proc}
\end{figure}
\subsubsection{Program} 
The \emph{Program} class is a subclass of \emph{Entity} that is used to  synthesize specific instances of \emph{Problem} and \emph{Processor} into an implementation of the AQO algorithm.  A \emph{Program} is the primary input to the \emph{Sapphire} simulation component and two different types can be constructed, physical or logical. The main difference between these two types for \emph{Program} is the presence or absence of a high-level logical \emph{Problem} definition. 
\begin{figure}[ht]
\begin{center}
\includegraphics[width=\textwidth]{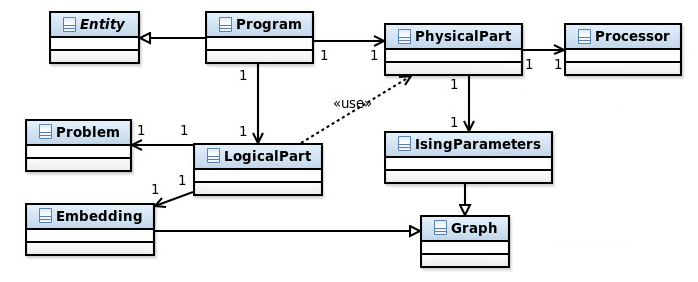}
\end{center}
\caption{The dependencies of the \emph{Program} class. The presence of a \emph{Problem} distinguishes a logical program from a physical program, while both class types have an associated \emph{Processor}. \label{fig:prog}}
\end{figure}
\par
As shown in Fig.~\ref{fig:prog}, type-switching is accomplished by composing \emph{Program} with two classes: \emph{Logical Part} and \emph{Physical Part}. The physical part of a \emph{Program} encapsulates the physical representation of the time-dependent Hamiltonian defined in Eq.~(\ref{eq:tdh}). This includes a reference to a \emph{Processor} and the parameters defining the final Ising Hamiltonian as well as the annealing schedule for each qubit. The logical part of a \emph{Program} encapsulates a physical program as well as a reference to the specified \emph{Problem} entity that is being solved. While the physical part of a \emph{Program} entity is always required, the logical part is not. For \emph{Analyst} use cases, the \emph{Program} always has a logical part. In the absence of a logical input, the \emph{Program} corresponds to an \emph{Engineer} defined instance of an Ising Hamiltonian.
\par
The mapping of the \emph{Logical Part} into the \emph{Physical Part} generates an \emph{Embedding} of the \emph{Problem} into the \emph{Processor}. As described in Sec.~\ref{sec:aqc}, embedding generates a map between each logical vertex and a connected subgraph in the \emph{Processor}. Within \emph{JadeD}, this is accomplished using a subclass of \emph{Graph} called \emph{Embedding}. The \emph{Embedding} class finds an embedding of the \emph{Logical Part} into the provided \emph{Processor} and \emph{Hardware}. The current \emph{Embedding} class supports the maximal minors methods described by Klymko et al. \cite{Klymko2014}. Its use is limited to a $K_{4,4}$, but the extensibility of \emph{Embedding} means that the additional, greedy methods described by Klymko et al.~can also be incorporated.
\subsection{NiCE}
\label{sec:nice}
The \emph{NiCE} component is responsible for accepting user input, returning JADE output, and managing the computational workflow. It also provides a graphical frontend for JADE. \emph{NiCE} is an existing open-source project that was leveraged for reducing development time and ensuring extensibility. In addition to I/O management, the \emph{NiCE} component orchestrates the interactions between the \emph{JadeD} and \emph{Sapphire} components. It enables users to create input files, launch simulations and examine program metrics. 
\par
\emph{NiCE} is based on a client-server model, where the server handles primary data management and the client acts as the user frontend. It is also possible for the server to manage remote workloads including, for example, simulations launched on remote hosts. We use the \emph{NiCE} server as the primary means for launching and monitoring numerical simulations on both local and remote machines.
\par
\begin{figure}[ht]
\begin{center}
\includegraphics[width=\textwidth]{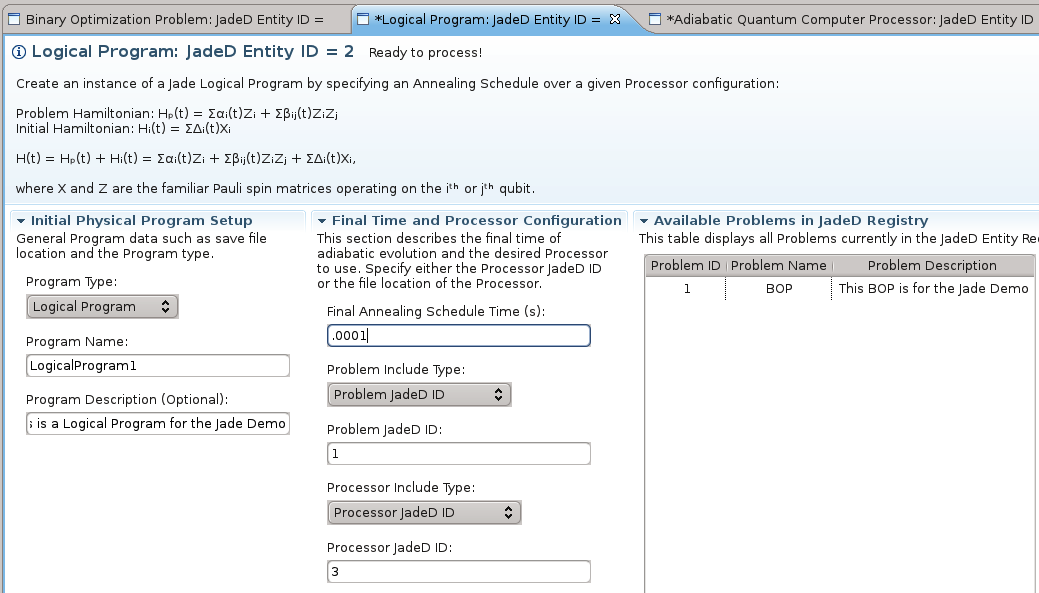}
\end{center}
\caption{A cropped screenshot from the \emph{NiCE} client for JADE showing the synthesis of a \emph{Program} from a logical \emph{Problem} and a selected \emph{Processor}.}
\label{fig:nice}
\end{figure}

We have developed several plug-ins for \emph{NiCE} that allow direct interaction with the \emph{JadeD} component for the creation and revision of the \emph{Problem}, \emph{Processor}, and \emph{Program} entities. A screenshot of one such \emph{NiCE} form is provided in Fig.~\ref{fig:nice}. \emph{NiCE} is based on the Open Source Gateway Initiative (OSGI) framework that, among other things, permits dynamic registration of services. We use \emph{NiCE}'s implementation of dynamic registration to recognize and load user-defined plug-ins into JADE. This feature permits, for example, user-defined methods for simulation that are developed independently from JADE to be added during runtime. Additional information about \emph{NiCE} is available from its website \cite{WWWNICE}.

\subsection{Sapphire}
\label{sec:sapphire}
\emph{Sapphire} is the JADE component responsible for profiling \emph{Program} entities. This includes carrying out numerical simulations of the quantum dynamics as well as other characterizations such as computing the time-dependent energy eigenspectra and computational error rates. While its primary use is to compute the \emph{Result} of a \emph{Program}, \emph{Sapphire} permits a robust set of possible use cases. This is a result of our use of a plug-in architecture to support user-defined extensions to \emph{Sapphire}. For example, numerical simulation techniques can be tailored to specific questions or physical assumptions. This promotes analysis at any desired fidelity and gives the user the ability to compare different simulation techniques against experimental benchmarks. 
\begin{figure}[h!]
\begin{center}
\includegraphics[width=\textwidth]{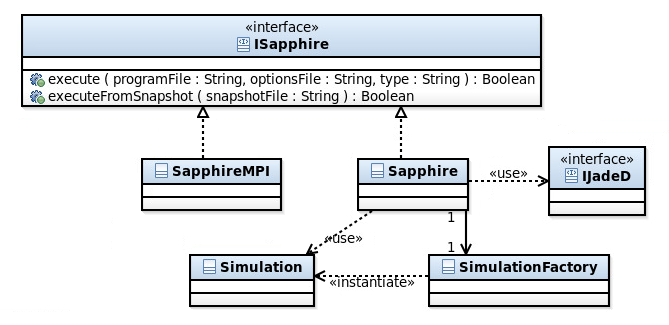}
\end{center}
\caption{The \emph{ISapphire} interface expresses both the \emph{Sapphire} and \emph{SapphireMPI} classes. \emph{Sapphire} makes use of the factory design pattern for generating \emph{Simulations} that are labeled by the \textit{type} argument to execute. \emph{SapphireMPI} has an identical structure. }
 \label{fig:sapphire}
\end{figure}
\par
The extensibility of \emph{Sapphire} is achieved through the interplay of a number abstractions and design patterns, as shown in Fig.~\ref{fig:sapphire}. \emph{Sapphire} only exposes a few methods to external clients through the \emph{ISapphire} interface. This decoupling between behavioral definition and actual implementation allows \emph{Sapphire} to take on a number of varied forms. For example, JADE currently provides a \emph{Sapphire} implementation for multi-threaded, shared memory architecture. We have also implemented \emph{SapphireMPI}, which uses the MPI (Message Passing Interface) library to execute simulations on distributed architectures. The most significant difference between the two implementations is the MPI dependency and the need to perform unique initialization steps for \emph{SapphireMPI} prior to beginning the numerical simulation.
\par
All implementations of \emph{Sapphire} must define the method \emph{execute}. When \emph{execute} is invoked, \emph{Sapphire} utilizes the \emph{JadeD} file-parsing capabilities to construct the \emph{Program} object defining the parameters of the numerical simulation. \emph{Sapphire} next parses the simulations options provided by the user to create a \emph{Simulation} object using the \emph{SimulationFactory}. The \emph{Simulation} class is the basis for the extensibility of \emph{Sapphire} using plug-in libraries. A plug-in is essentially a subclass of \emph{Simulation} that provides a specialized numerical or algorithmic approach to simulation.
\subsection{Simulation Plug-ins}
\label{sec:sim}
The \emph{Simulation} class is the primary unit of functionality within \emph{Sapphire} and it is used to encapsulate a specific mathematical evolution of a quantum state. The factory design pattern allows \emph{Sapphire} to remain completely agnostic to simulation details. However, there is a specific sequence of execution statements that are a necessary part of \emph{Sapphire}. Program execution always begins with an initialization statement followed by a loop over a time-dependent solver. Once the exit condition is met, i.e., when $t=T$, the computational state undergoes readout before the program issues finalization commands. All plug-ins for \emph{Sapphire} must adhere to the \emph{Simulation} class functionality defined below.
\begin{itemize}
\item \emph{initialize}: This method is used primarily to initialize quantum state of the simulation. Additional tasks include setting up any pre-simulation conditions or parameters. 
\item \emph{anneal}: This method is called every time step by \emph{Sapphire} to advance the program quantum state. Developers should implement this method to update the state vector with the mathematics inherent to a specific technique for solving the time-dependent Schr\"{o}dinger equation. 
\item \emph{queryState}: This method is used to query the state of the simulation, including the computational state of the simulated program. The output generated by this method is highly variable and it can include the internal representation of the quantum computational state or the complete eigenenergy spectrum written to an output file. These output files can also be used as checkpoints for restarting the simulation.
\item \emph{measure}: This method is called after \emph{anneal} completes and it represents measurement of the final computational state. 
\item \emph{finalize}: This method is used for any final calculations or clean up routines. 
\end{itemize}
Developers of simulation plug-ins must subclass \emph{Simulation} and implement the purely virtual \emph{anneal} method. All other methods have default implementations that can be overwritten for specialized functionality. JADE also provides a specialized \emph{HamiltonianGenerator} abstraction that permits decoupling of numerical dynamics from the actual form of the Hamiltonian describing the system. 
\subsubsection{Plug-in Examples}
\label{sec:plugins}
The \emph{Sapphire} plug-in architecture maintains extensibility to new simulation methodologies. A plug-in represents a user-created library that implements the \emph{Simulation} class defined above. JADE users are therefore able to tailor quantum computing simulation techniques to specific problems or metrics of interest. We provide examples of plug-ins that implement \emph{Simulation} below. 
\begin{figure}[ht]
\begin{center}
\includegraphics[width=\textwidth]{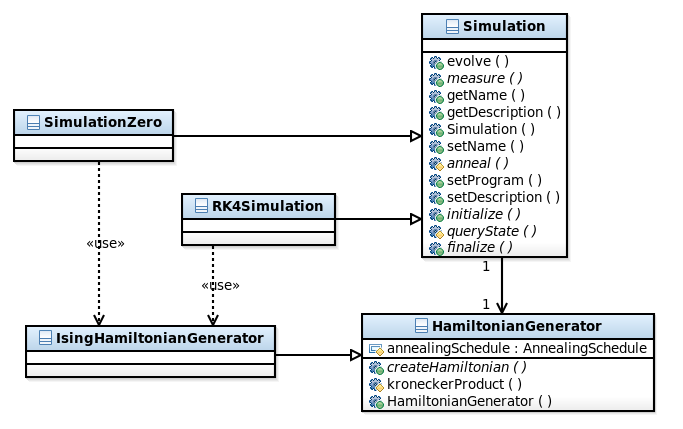}
\end{center}
\caption{Examples of the plug-ins that implement the \emph{Simulation} class. SimulationZero and RK4Simulation both implement subclass \emph{Simulation} and also make use of IsingHamiltonianGenerator, a subclass of the \emph{HamiltonianGenerator}. \label{fig:plugins}}
\end{figure}
\begin{itemize}
\item \emph{SimulationZero}: This plug-in provides a zero-th order approximation about the state of the computational register. Specifically, this simulation calculates the time-dependent eigenspectrum and instantaneous eigenstates of the time-dependent Hamiltonian defined by a \emph{Program}. SimulationZero does not provide information about the quantum dynamics but essentially diagonalizes the Hamiltonian at each time step. This analysis provides information about the time-dependent energy gap. Our implementation makes use of the Eigen library, which is an open-source C++ template library for linear algebra \cite{WWWEIGEN}. 
\item \emph{RK4Simulation}: This plug-in provides a fourth-order Runge-Kutta solver for the time-dependent Schr\"{o}dinger equation as in Eq.~(\ref{eq:tdse}).  RK4Simulation uses two time steps, one for the outer \emph{anneal} method which updates the Hamiltonian and a second for the inner \emph{evolve} loop that numerically solves a finite-difference equation. For each \emph{evolve} time step, the plug-in updates the quantum state and for each \emph{anneal} it computes the instantaneous eigenspectrum.  The plug-in also implements the \emph{queryState} method to provide a \emph{Snapshot} output that contains details about the computational state and eigenspectrum. Simulation options include the time steps, number of \emph{Snapshot} files created, and number of eigenstates reported by \emph{queryState}. This plug-in also makes use of the linear algebra functionality provided by the Eigen library. 
\item \emph{FOPSimulation}: The FOPSimulation plug-in is based on a first-order perturbative solution to the time-dependent Schr\"{o}dinger equation. It evolves a pure state according to a first-order Magnus expansion for the time-dependent propagation operator. Numerically, the propagation operator is diagonalized by the \emph{anneal} method and applied successively to the state during the \emph{evolve} method. This method has an error of $\mathcal{O}(\Delta t^3)$. Similar to the other simulation methods, Eigen is used to perform the matrix exponential and matrix-vector multiplications.
\end{itemize}
\subsection{Testing Framework}
\label{sec:jvv}
\begin{figure}[h]
\begin{center}
\includegraphics[width=5.0in]{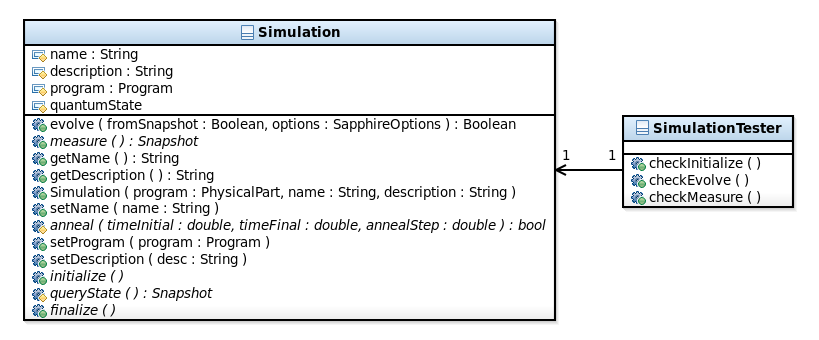}
\end{center}
\caption{\emph{SimulationTester} is external to \emph{Simulation} but capable of accessing its methods. }
\label{fig:st}
\end{figure}

The design and implementation of JADE relies heavily on test-driven development. A formal and rigorous testing model was defined before any actual product code was developed. This has ensured that (1) the functionality of each test unit was defined prior to its implementation and (2) the implementation of each source unit was fully compliant with the predetermined functionality. We employed test-driven development by modeling and designing surrogate classes whose sole purpose was for unit testing critical behavior in actual JADE classes. An  example is shown in Fig.~\ref{fig:st}, where we test the \emph{Simulation} class using surrogates for most objects in the Sapphire component. There is a corresponding \emph{SimulationTester} class. Every class in JADE has a corresponding test class in order to provide the greatest assurance that the code adheres to design requirements.

\section{Usage Example}
\label{sec:results}
\begin{figure}[ht]
\begin{center}
\includegraphics[width=.4\textwidth]{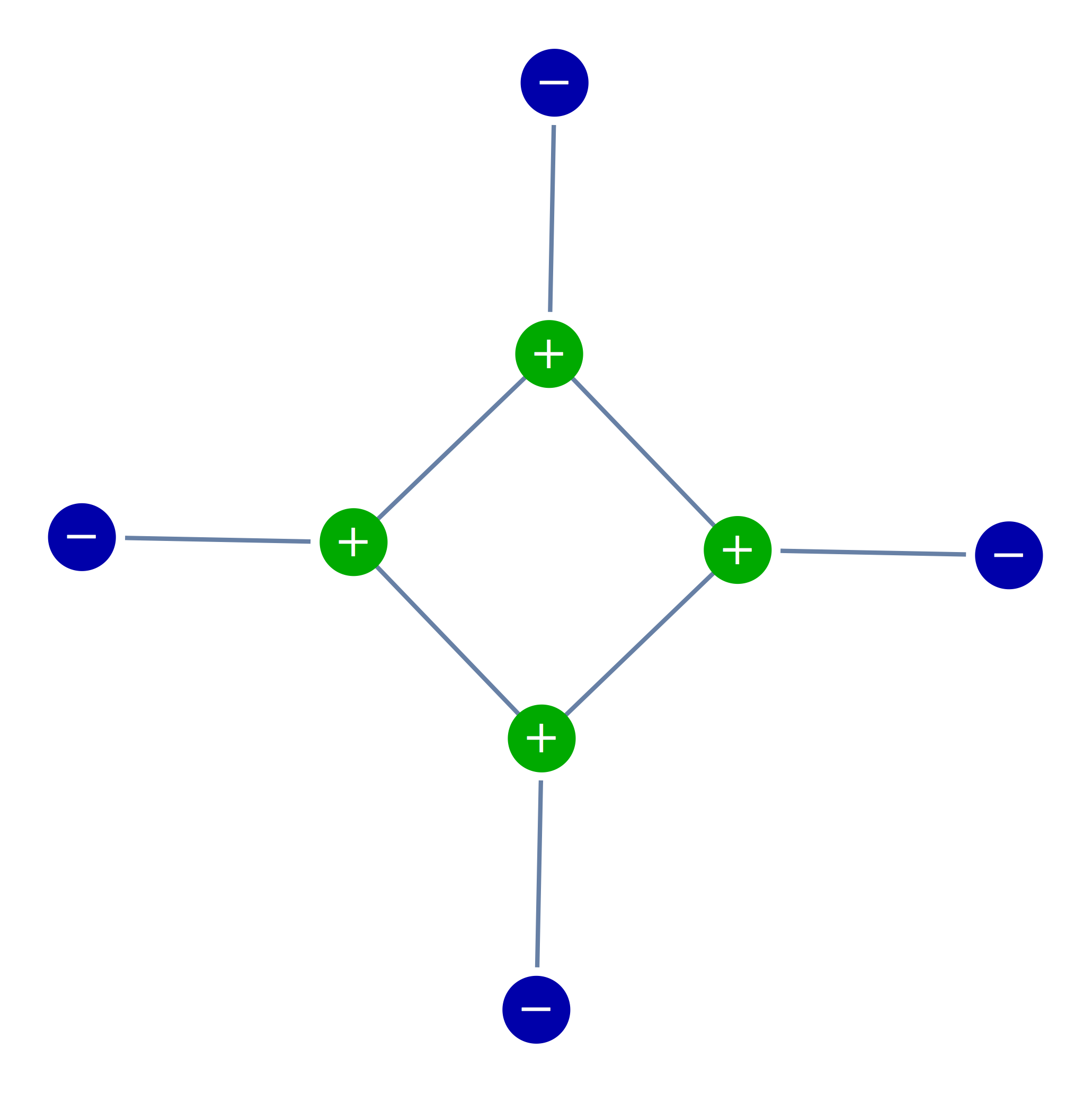}
\end{center}
\caption{\small \sl The graphical representation of the 8-qubit Ising model investigated by Boixo et al.\cite{Boixo2013}. Vertices 1-4 (green) represent biases of $+1$ and vertices 5-8 represent biases of $-1$. All the edges represent $+1$ couplings between connected vertices. \label{fig:8qubit}}
\end{figure}

As an example of how JADE can be used for evaluating quantum programs, we present results based on the recent experimental  benchmarks reported by Boixo et al. \cite{Boixo2013}. Their work was performed on the Rainier processor from D-Wave Systems, Inc. and used the 8-qubit Ising model represented in Fig.~\ref{fig:8qubit}.
\par
Boixo et al. showed both theoretically and experimentally that the 8-qubit model in Fig.~\ref{fig:8qubit} exhibits a unique behavior. This particular 8-qubit problem exhibited a distinctive behavior that differentiates between the quantum and classical annealing dynamics. The Ising Hamiltonian has a 17-fold degenerate ground state. They used multiple runs of the developed program on the Rainier processor to recover all 17 ground states from computational readout. 
\par
We have used the benchmark developed by Boixo et al.~to demonstrate the functionality of JADE. Specifically, we defined an 8-qubit \emph{Processor} supporting the $K_{4,4,}$ (bipartite) connectivity familiar from the unit cell in the Rainier processor as shown in Fig.~\ref{fig:dw1hw}. We used an \emph{Embedding} entity based on the maximal minor method discussed by Klymko et al. \cite{Klymko2014} and we matched the mapping taken by Boixo et al. We programmed linear annealing schedules, i.e., $A(t) = t/T$ and $B(t) = 1 - t/T$, and a final time of $T = 30\tau$, where $\tau = h/ E_0$ defines time relative to the energy scaling $E_0$ of the Hamiltonian $H(t)$, i.e., $H(t) \rightarrow H(t)/E_0$. Fortuitously, the value of $E_0$ drops out of these calculations as we measure time relative to $T$, i.e., as $t/T$. We have neglected constraints on the controls, as the Ising parameters were very simple, and we have neglected all forms of noise in the hardware physics.
\par
The developed \emph{Program} entity was then simulated using the RKSimulation plug-in described in Sec.~\ref{sec:plugins}. The simulation options given to this fourth-order Runge-Kutta finite-difference solver invoked a quasi-static approximation for the Hamiltonian. That is to say, we used an \emph{evolve} time step of 0.0001 $\tau$ with updates to the Hamiltonian made during every \emph{anneal} with a time step of 0.05 $\tau$. The computational registers were initialized to the exact ground state of the initial Hamiltonian in Eq.~(\ref{eq:HI}).  For diagnostics, we computed the complete eigenspectrum every 3 $\tau$ and output both the spectrum and the complete quantum state as part of a \emph{Snapshot}. The \emph{measure} method returned an ordered listing of the output states with their associated probabilities in the generated \emph{Result} entity.
\begin{figure}[ht]
\centering
\begin{subfigure}
\centering
\includegraphics[width=0.45\textwidth]{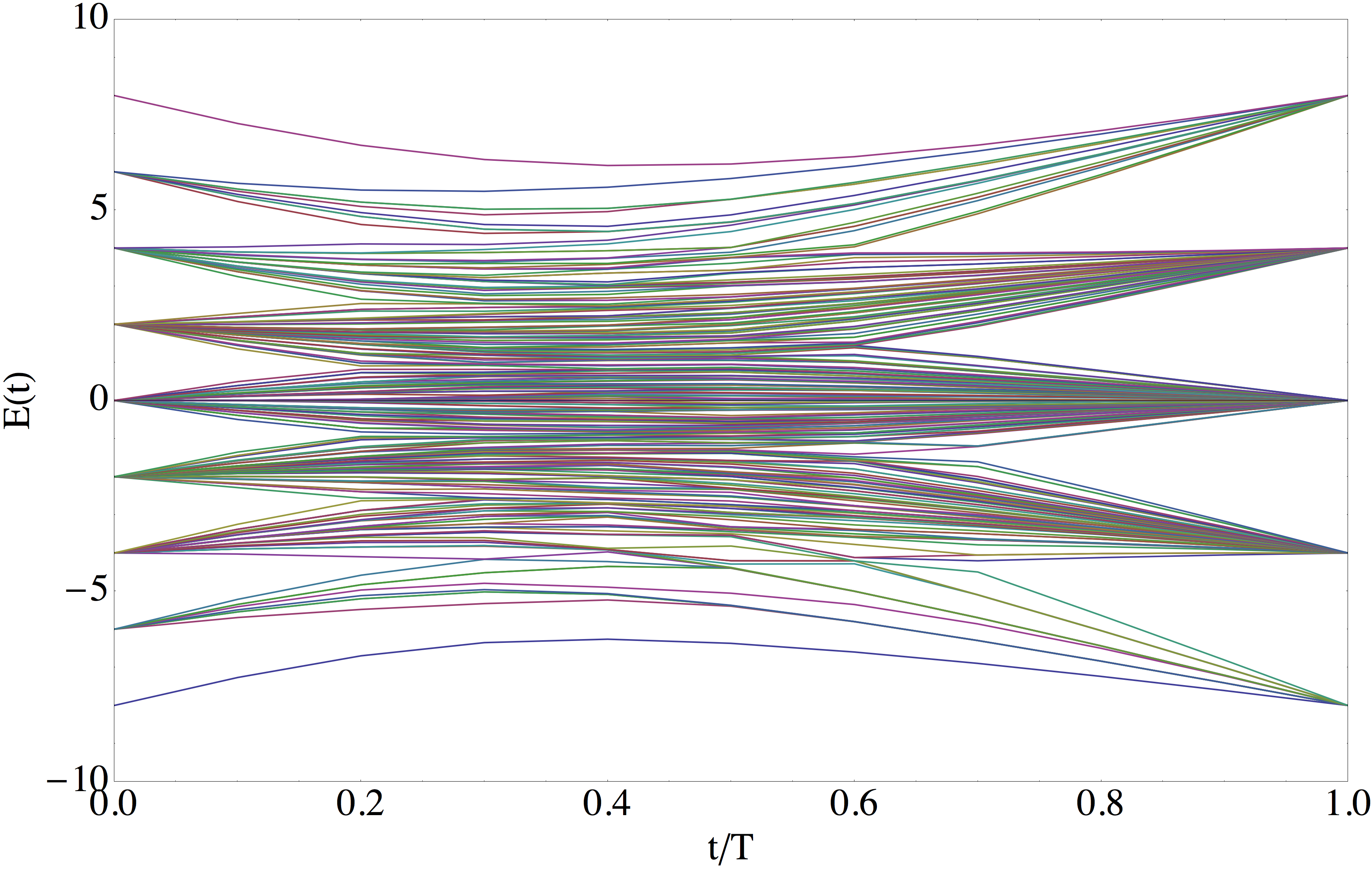}
\end{subfigure}
\begin{subfigure}
\centering
\includegraphics[width=0.45\textwidth]{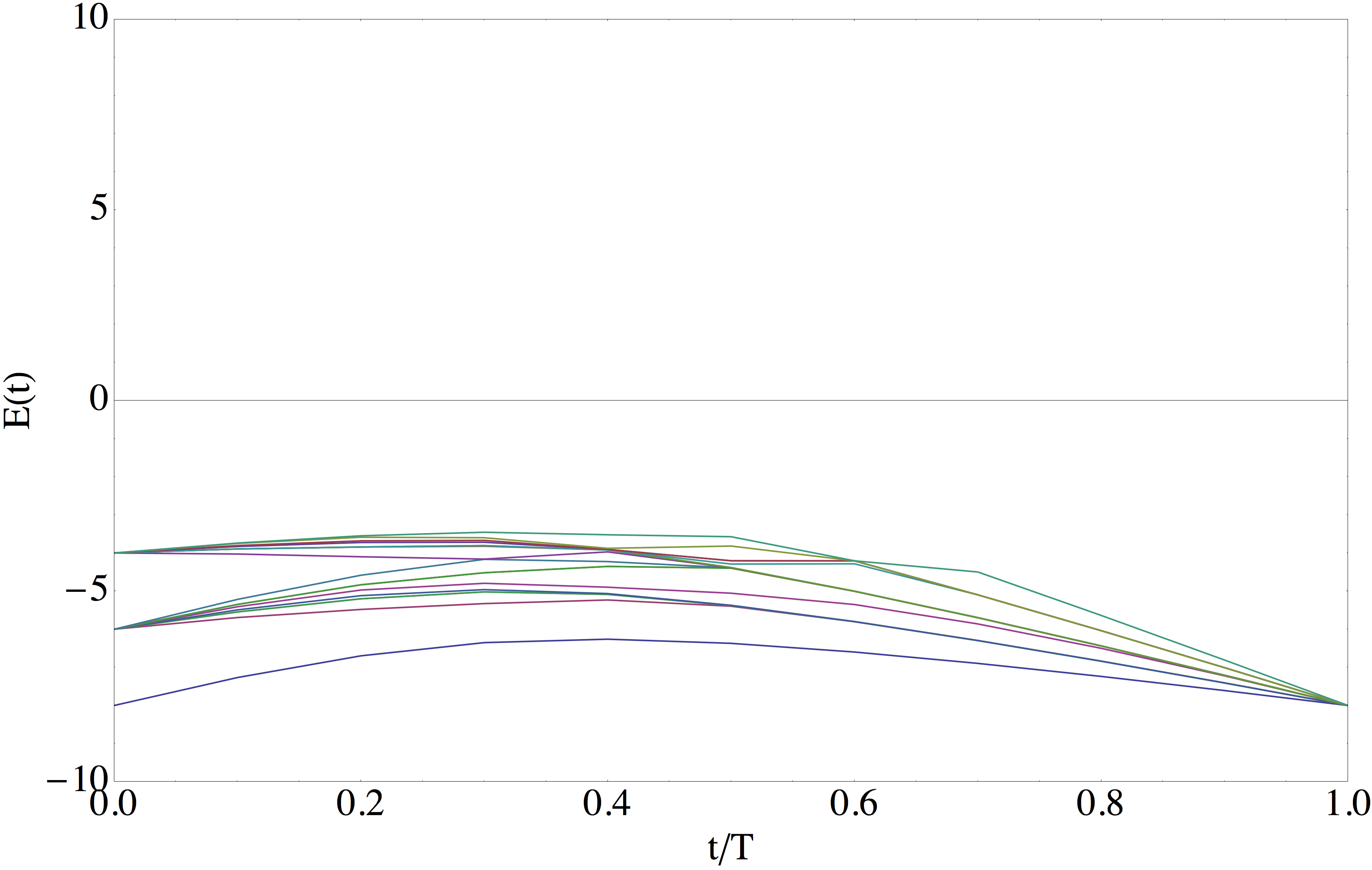}
\end{subfigure}
\caption{(left) The complete time-dependent eigenspectrum of the 8-qubit program.  (right) Time-dependent spectrum for those states terminating in computational ground states. Spectra are computed every 3 $\tau$ for a total of 11 points for each of the 256 spectra lines. \label{fig:groundspectrum}}
\end{figure}
\par
The complete time-dependent eigenspectrum computed by JADE is shown in the left panel of Fig.~\ref{fig:groundspectrum}. This consists of $2^8$ = 256 lines representing the time-dependent energies of the 256 eigenstates of the Hamiltonian. At the final time $T$, there are 17 ground states  with eigenenergy $-8$. This matches the eigenenergy and degeneracy derived by Boixo et al. The 17 time-dependent spectra that result in a ground state at the final time are shown in the right panel of Fig.~\ref{fig:groundspectrum}. The presence of kinks in the plot indicate that several states undergo avoided crossings with higher energy levels. Recall that the definition of the spectral gap $\Delta(t)$ in Eq.~(\ref{eq:gap}) did not distinguish between those instantaneous excited states that terminate in the final ground state manifold from those excited states that remain excited at time $T$. States terminating in the ground state manifold are not computational errors, but transitions from those states to higher lying excited states can contribute to the observed error rate. 
\par
The computed populations for the 17 ground states at time $T$ are presented in Table 1 alongside the corresponding computational basis state. It is evident that the first 16 states, i.e., the manifold of states with qubits 1-4 in the 0 (spin down) state, have approximately equal probability while the 17-th state is roughly two orders of magnitude less. However, all the ground states are significantly more likely than the 18-th most probable state, which has a probability much less than $10^{-6}$.
\begin{table}
\caption{Degenerate ground states of the 8-qubit model and their computed probabilities. \label{tab:pop}}
\begin{center}
\begin{tabular}{|r|c|r|}
\hline 
Decimal & Binary & Probability \\
\hline \hline
0  & 0000 0000 & 0.0582245  \\ \hline
1  & 0000 0001 & 0.0598409  \\ \hline
2  & 0000 0010 & 0.0598409  \\ \hline
3  & 0000 0011 & 0.0620211  \\ \hline
4  & 0000 0100 & 0.0598409  \\ \hline
5  & 0000 0101 & 0.0627384  \\ \hline
6  & 0000 0110 & 0.0620211  \\ \hline
7  & 0000 0111 & 0.0651488  \\ \hline
8  & 0000 1000 & 0.0598409  \\  \hline
9  & 0000 1001 & 0.0620211  \\ \hline
10  & 0000 1010 & 0.0627384  \\ \hline
11  & 0000 1011 & 0.0651488  \\ \hline
12  & 0000 1100 & 0.0620211  \\ \hline
13  & 0000 1101 & 0.0651488  \\ \hline
14  & 0000 1110 & 0.0651488  \\ \hline
15  & 0000 1111 & 0.0677486  \\ \hline
255  &1111 1111  & $4.79745 \times 10^{-4}$ \\ \hline
\end{tabular}
\end{center}
\end{table}
\begin{figure}[ht]
\begin{center}
\includegraphics[height=2.73in]{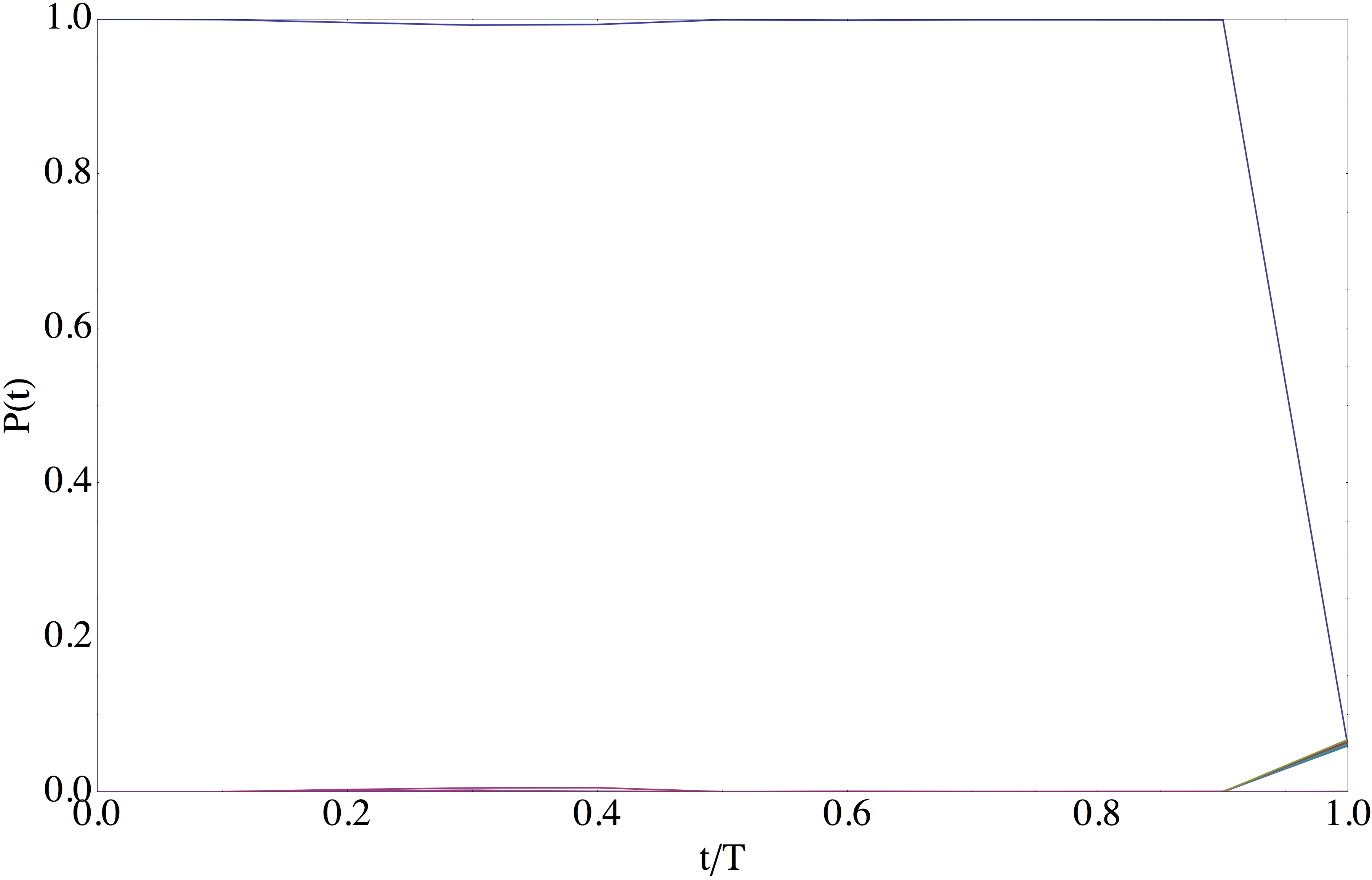}
\end{center}
\caption{\small \sl Time-dependence of the population in the computational basis. The resolution is 11 points over the the range $[0, T]$.  \label{fig:groundpop}}
\end{figure}
\par
The time-dependence of the instantaneous population in the computational basis is shown in Fig.~\ref{fig:groundpop}. Recall that the system is initialized in the singular computational ground state, as indicated by maximum probability at time $t=0$. As time progresses, the population remains in the instantaneous ground state until $t \approx 0.9 T$ At this point in the program schedule, the energy gap between the ground state and the lowest lying excited states has narrowed sufficiently to permit population transfer, thereby violating the adiabatic condition. At this point in the dynamics, however, the lowest-lying excited states represent instantaneous states that will terminate in the ground state at time $t=T$. There are 16 such states participating in the apparent convergence to approximately 15/16 of the total probability and, as shown in Table \ref{tab:pop}. The 17-th ground state is not visible in this plot, due to the scale of its contribution, however it undergoes a similar behavior and contains approximately $1/16^2$ of the population. Approximately $15/256$ of probability is distributed over the remaining $239$ excited states.
\par
Our simulation of the 8-qubit program appears to be in qualitative agreement with the experimental and theoretical results of Boixo et al. \cite{Boixo2013}. However, there are several key differences between their program and ours. First, the annealing schedules used by Boixo et al. are not linear and we expect that impacts our comparison of observed and computed probabilities. Second, we have not incorporated any sources of noise into our simulation studies, whereas previous experiments on the D-Wave processors have suggested influences of thermal noise may be significant. Nevertheless, our intention of this demonstration has been to provide a verifiable example that JADE is useful for developing quantum programs and supporting benchmark analysis.

\section{Discussion}
\label{sec:discussion}
The present availability and continuing development of adiabatic quantum computing hardware opens up new avenues of research for defining methods of quantum programming and computational benchmarking. Experimental studies are necessary for measuring  actual computational power of processors and for improving programming practices. Test vectors appropriate for benchmark studies must be well defined and the associated difficulty well understood  in order to reliably measure the influence of programming and processor methodologies. 
\par
Our contribution has been to develop a software environment that offers an interactive approach to programming the adiabatic quantum optimization algorithm. JADE parametrizes the process of programming the AQO algorithm and it offers opportunities for tuning each step. JADE, or software like it, is needed for standardizing program studies as well as for optimizing program performance. In particular, we have shown how there are many tunable parameters that contribute to the implementation of the AQO algorithm in a processor modeled by a spin-glass system. JADE offers opportunities for optimizing performance across program parameters by exposing these interfaces to the user. Similarly, the \emph{Program} entity introduced here is one example of a data structure that captures program instance and, consequently, standardizes program specification.  We have used \emph{Program} to initialize numerical simulations, but it would also be possible to submit these program directly to the D-Wave Systems processor.  The direct interaction between JADE and the underlying hardware is currently under investigation.
\par
JADE also provides a plug-in architecture to enable extensions to functionality through user-defined programming, simulation, and diagnostic methodologies. We have discussed our implementation of two high-level logical input methods (BOP and QUBO problems), reconfigurable processor definitions in terms of hardware size and connectivity), and multiple numerical simulation methods for computing the time-dependent eigenspectra and eigenstates. The extensible design of JADE permits each of these features to be easily replaced by newer and potentially more versatile methods without revision to the existing code base.
\par
Finally, the programming sequence for the AQO algorithm summarized in Fig.~\ref{fig:flowchart} is sufficient for the currently available hardware models. However, we anticipate that future hardware and programming models will modify the steps taken in compiling an adiabatic quantum algorithm down to a (future) quantum processor. In particular, our approach does not account for fault-tolerance, quantum error correction, or quantum control techniques, which are expected to be useful for the broader AQC paradigm \cite{Young2013}. Nevertheless, we believe JADE exemplifies the type of programming environment currently needed by the quantum computer science community for evaluating the performance of current and future quantum processors.

\subsection*{Acknowledgments}
This work was supported by the Lockheed Martin Open Innovation Program at Oak Ridge National Laboratory. The authors thank Greg Tallant and Peter Stanfill of the Lockheed Martin Aeronautics Division for technical assistance. H.~S.~thanks the Department of Energy Science Undergraduate Laboratory Internship (SULI) program. T.~S.~H.~thanks Owen S. Humble for technical discussions. This manuscript has been authored by a contractor of the U.S. Government under Contract No.~DE-AC05-00OR22725. Accordingly, the U.S. Government retains a non-exclusive, royalty-free license to publish or reproduce the published form of this contribution, or allow others to do so, for U.S. Government purposes. Developers interested in licensing JADE should contact the authors.

\section*{References}
\providecommand{\newblock}{}

\end{document}